# Modeling the production of VCV sequences via the inversion of a biomechanical model of the tongue.

, *Pascal Perrier[1], Liang Ma[1,2] & Yohan Payan[3]*.

[1]Institut de la Communication Parlée, UMR CNRS 5009, I.N.P. Grenoble, France
[2]Laboratoire Parole et Langage, UMR CNRS 6057, Univ. de Provence, Aix-en-Provence, France
[3]TIMC, UMR CNRS 5009, I.N.P. Grenoble, France

perrier@icp.inpg.fr, maliang@icp.inpg.fr, Yohan.Payan@imag.fr

## Abstract

A control model of the production of VCV sequences is presented, which consists in three main parts: a static forward model of the relations between motor commands and acoustic properties; the specification of targets in the perceptual space; a planning procedure based on optimization principles. Examples of simulations generated with this model illustrate how it can be used to assess theories and models of coarticulation in speech.

## 1. Introduction

Modeling the production of speech sequences from the phonemes to the acoustic signal is an old but still very challenging issue in speech communication research. There are two major problems when dealing with this issue. First, for a given phoneme, depending on the phonetic context, on the speaking style or on the speaking rate, various physical characteristics are observed [1]. Second, in order to produce a given spectral characteristics of the acoustic signal, many articulatory configurations can theoretically be selected [2].

Numerous studies of the literature have addressed the *many-to-one* issue of the articulatory-to acoustic relations. In the context of speech sequences, the problem can often be regularized first by specifying the whole desired time-to-time trajectory in the acoustic space and, second, by searching for an articulatory trajectory that is continuous and smooth. Continuity and smoothness can be obtained thanks to the construction of *continuity maps* that limit the time-to-time variation of the commands [3] or to the minimization of a certain *cost* over the whole sequence.[4]. In order to account for the variability of the physical realizations of a phoneme, the desired trajectory can connect target regions in the perceptual space instead of target points [5].

These different studies are highly interesting and contribute largely to the development of speech inversion techniques and to the understanding of speech motor control. However, they have two main drawbacks. First, they require the specification of the whole time-to-time desired trajectory in the perceptual space. Second, they give no account of the impact of the intrinsic physical properties of the speech apparatus onto the trajectory in the perceptual space.

In the line of Dang & Honda [6] who control their physiological model on the basis of targets without specifying the whole trajectory, or of Zandipour *et al* [7] who looked for simplified specifications of the desired trajectory that could take into account the influence of biomechanics, we propose a target based control model, accounting for sequence planning, and for the interaction between motor commands, timing and physical properties of the articulators. It consists of three main parts: a static forward model of the relations between motor commands and acoustic properties; the specification of targets in the perceptual space; a planning procedure based on optimization principles.

## 2. Static Forward Model

We call *static forward model* a functional model of the relations between the motor commands sent to the tongue muscles and the spectral characteristics of the acoustic speech signal. It was elaborated using a 2D biomechanical model of the tongue.

### 2.1. The 2D biomechanical tongue model [8, 9]

Elastic properties of tongue tissues are accounted for by finite-element (FE) modeling. Muscles are modeled as force generators that (1) act on anatomically specified sets of nodes of the FE structure, and (2) modify the stiffness of specific elements of the model to account for muscle contractions within tongue tissues. Curves representing the contours of the lips, palate and pharynx in the midsagittal plane are added to specify the limits of the vocal tract. The jaw and the hyoid bone are represented in this plane by static rigid structures to which the tongue is attached. Changes in jaw height can be simulated through a single parameter that modifies the vertical position of the whole FE structure. The model is controlled according to the $\lambda$ model [10] that specifies for each muscle a threshold length, $\lambda$, where active forces start. If the muscle length is larger than $\lambda$, muscle force increases exponentially with the difference between the two lengths. Otherwise there is no active force. Hence, muscle forces are typically non linear functions of muscle lengths. The control space is called the $\lambda$ space (for more details see [8, 9])

### 2.2. Describing the relations between motor control variables and formant patterns

It was shown that the tongue model gives a fairly realistic account of the impact of muscle recruitment on the tongue shape, and of the variety of tongue deformations and displacements during speech production [8,9,11]. Hence, the model was used in order to obtain a comprehensive and realistic description of the relations between motor control variables, tongue shapes and spectral characteristics.

*2.2.1. From motor commands to tongue shapes*

Around 8800 different simulations were generated, which describe a large variety of tongue shapes; including consonantal ones. To do so, the λ space was randomly sampled according to a uniform distribution, which range of variation was adapted for each muscle in order to get a fair account of the extreme tongue shapes observed during speech production. Jaw position was the same in all cases. Starting from the rest position, movements were simulated toward each of the 8800 tongue shapes, by shifting the control variables slowly and at a constant rate of shift with the following timing:

- Transition Duration from rest position to the selected tongue shape: 50 ms
- Hold duration of the final configuration: 150 ms.

This timing was chosen, in order to make sure that at the end of the 8800 simulations, the tongue was static, that no target undershoot occurred, and that the final tongue shape was a true image of the motor commands..

*2.2.2. From tongue shapes to formants*

Formant values were computed for the final tongue shape of each simulation. In this aim, the area function was computed with an enhanced version of Perrier et al.'s model [12]. The biomechanical model does not include any description of the lips. Hence, for each tongue position, two area functions were generated for two different lip areas corresponding to average lip areas respectively for spread lips (3 cm$^2$) and for rounded lips (0.5 cm$^2$).

The first three maxima of the vocal tract frequency response were calculated with a harmonic model of the vocal tract taking into account viscosity and wall vibration losses as well as lip radiation impedance [13].

*2.2.3. Results*

Figure 1 shows the dispersion of the positions of the 17 nodes determining the upper tongue contour in the FE structure. It can been seen that the set of simulations includes open vocal tract configurations as well as occlusions in the velo-palatal and alveolar regions.. The jaw position used for all simulations is shown by the constant position of the lower incisor.

Figure 2 depicts the formant distributions in the (F1, F2) plane for the two different lip areas. Because of the non-linearity of the relations between the articulatory and the acoustic domains, the distributions are not uniform any more, and the majority of the configurations are located for spread lips along the [i, e, ε, a] axis and for rounded lips along the [y, o] direction.

**2.3. Learning the relations between motor control variables and formant patterns**

From these two data sets, a functional model of the relations between motor control variables (λ) and (F1,F2,F3) formant patterns has been elaborated. We call it *static forward model*. Indeed, in the line of Jordan [14] or Kawato *et al.* [15] it is hypothesized that this model could correspond to an internal representation (or *forward model*) that the Central Nervous System could use to determine the motor commands for a given perceptual effect. In the current form of the control model, this model is static, because it does not take into account information about the dynamics of the motor system. It should also be noted that, since only two lip areas were considered to generate formant patterns, lip control could not be learned from these simulations. Hence, two models were elaborated respectively for the rounded and for the spread lips. In this paper, only the spread lips case will be considered.

We have to model a many-to-one and non-linear relation. For this kind of problem, neural networks based on Radial Basis Functions (RBF) have been proved to be very powerful [16, 5]. Hence, using the MATLAB Neural Networks toolbox, a neural network was trained on a subset of our data including 6400 randomly selected configurations. In this way, an optimal neural network was elaborated, using 400 Gaussian RBF. The accuracy of the model was tested on the remaining 2400 configurations that were not used to train the neural network, and the mean square error on the (F1, F2, F3) patterns was less than 3%.

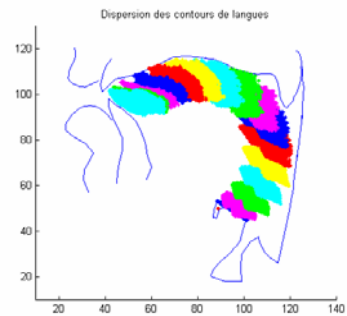

*Figure 1*: Dispersion of the 17 nodes positions located on the upper tongue contour in the mid-sagittal plane..

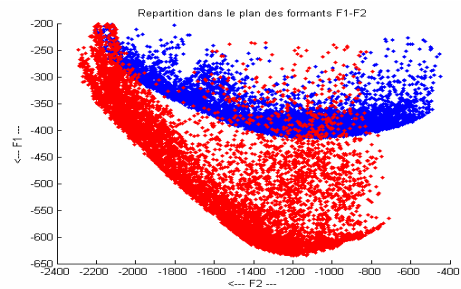

*Figure 2*: Dispersion in the (F1, F2) plane for a 3 cm$^2$ (light gray dots) and a 0.5 cm$^2$ lip area (dark grey dots)

## 3. Inferring motor commands from phonemes specification: inversion and optimization

The *static forward model* was used to infer from a sequence of phonemes an appropriate sequence of motor commands. In this aim, we defined, for each phoneme, correlates in the (F1,F2,F3) domain, and elaborated an inversion procedure to recover motor commands from the formant patterns.

### 3.1. Correlates of phonemes in the formant space.

In this study, the phonemes inventory is limited to vowels [i, e, ɛ, a, œ, ɔ] and to the velar stops [k, g]. It is hypothesized that correlates can be found in the formant space both for vowels and for consonants, and that these correlates take the form of phoneme specific *target regions* in the (F1,F2,F3) space.. Target regions are supposed to contain all the acoustil realizations of a phoneme that are strictly equivalent from a perceptual point of view. In the current state of our work, these regions are defined as dispersion ellipses in the (F1,F2) and (F2,F3) planes, which are fully determined by their center coordinates (Fc1, Fc2, Fc3) and by the standard deviations (σF1, σF2, σF3).

For vowels, the specification of these regions was largely based on Calliope [18]. For stops, we used a kind of adaptation of the concept of *locus* [19], which we consider to be an ideal, non reachable, resonance frequency patterns of the vocal tract. To obtain an approximation of these ideal patterns for the velar consonants [k, g], we calculated the formants associated with the vocal tract configurations having an occlusion in the velar region. Table 1 summarize the values that were used for the different phonemes considered in this study.

*Table 1*: Values in Hz of the centers and the standard deviations defining the target regions of the phonemes

|  | [i] | [e] | [ɛ] | [a] | [œ] | [ɔ] | [k] |
|---|---|---|---|---|---|---|---|
| $F_{c1}$ | 288 | 375 | 503 | 629 | 498 | 550 | 251 |
| $F_{c2}$ | 2178 | 2097 | 1696 | 1296 | 1508 | 1019 | 1191 |
| $F_{c3}$ | 3005 | 2413 | 2782 | 2615 | 2566 | 2499 | 2629 |
| $\sigma_{F1}$ | 15 | 30 | 40 | 40 | 40 | 30 | 35 |
| $\sigma_{F2}$ | 120 | 70 | 80 | 130 | 120 | 60 | 350 |
| $\sigma_{F3}$ | 140 | 90 | 100 | 130 | 100 | 100 | 200 |

### 3.2. Inversion of the static forward model

Thus, many perceptually equivalent formant patterns are associated with each phoneme. Since many motor control variables are also associated with each formant pattern, numerous motor command sequences are likely to produce a given sequence of phonemes. To select one specific sequence among all the possible ones, additional constraints have to be considered. In this aim, we defined a specific cost, called C, which minimization allows the selection of an appropriate sequence of motor commands. C is defined as the sum of a *speaker-oriented* and a *listener-oriented* criterion [17].

The speaker-oriented criterion consists in the length of the global path that connects in the λ space the successive target commands associated with each phoneme of the sequence. The minimization of this criterion allows to keep the variation of the motor commands as low as possible. It is a smoothing technique in the motor command space, and it only concerns target commands.

The listener oriented criterion aims at ensuring that the intended perceptual effect will be produced for each phoneme of the sequence, in spite of the speaker oriented optimization. In this main, we defined a function that has a very low value for all the configurations within the dispersion ellipsis characterizing the phoneme, and that increases abruptly as soon as the formant patterns go out of the ellipsis.

## 4. Generation of VCV sequences: confronting models of coarticulation

The proposed model of motor commands planning from the specification of the phoneme sequences offers an interesting framework to test quantitatively various models of coarticulation in speech. Indeed, the planning can be tested on different kinds and different lengths of sequences, different proposals can also be tested for the time variation of the motor commands from a target command to the next; in addition, the biomechanical and dynamical properties of the tongue model permit to evaluate the specific impact of the physics on the measured variability associated with coarticulation [20].

To illustrate this statement we present examples of simulated tongue movements for [$V_1kV_2$] sequences obtained for two different models of coarticulation. The first model, called *sequential model*, minimizes the cost C for the whole sequence. Movement is then generated by shifting the motor control variables from a target to the next at a constant rate. The second model is inspired from Öhman's model [21], where the sequence of vowels is assumed to be the basis of speech movements and the consonants are seen as local perturbations of the vowel-to-vowel movement. In our "Öhman-like" model the planning of the target motor commands takes into account only vowels. The consonant target is the same for all kinds of vowel combination. To generate the movement, the targets are shifted at a constant rate from a vowel target to the next. The movement toward the consonant is generated by perturbing these linear shifts, at a specific time and for a specific duration.

Figures 3 and 4 show the displacement and the tangential velocity of a node located on tongue dorsum in the velar region for the sequences [*rest*-œ-e-k-i] and [*rest*-œ-e-k-a].in the two different modeling frameworks. In the sequential model, it can be seen that the horizontal position of [k] is slightly more anterior in the [i] context, and that velocity profiles start to differ only when the movement toward the consonant starts. In the other case, the consonant position is the same in both vowel contexts, but vowel [e] is slightly more anterior and higher in the [i] context. As a consequence, differences can be noted on the tangential velocity as early as the beginning of the [œ-e] transition. Finally, it should also be noted that the velocity profile of the movement toward the consonant has two peaks when the sequential model is used.

Comparing such results with articulatory data collected on real speakers, offers a good way to test quantitatively hypotheses about coarticulation.

## 5. Conclusion

We have developed a control model of the production of VCV sequences based on the concept of target regions in the perceptual space, and using a forward model and optimization techniques to infer optimal motor commands from the phonemic input. We have shown that it its current state, the model represents a good framework to further test hypotheses about speech production and speech coarticulation.

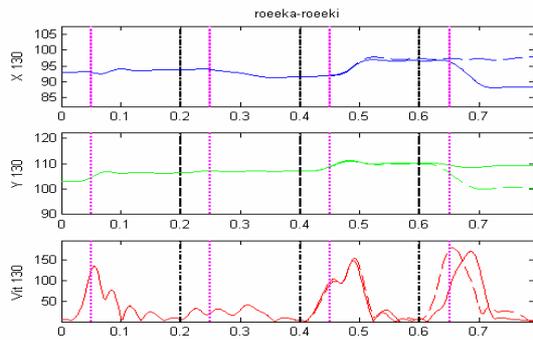

*Figure 3*: Sequential coarticulation model. Synthesized horizontal (top panel) and vertical (middle panel) displacements, as well as tangential velocity of a node located on tongue dorsum for the sequences [*rest*-œ-e-k-i] (solid line) and [*rest*-œ-e-k-a] (dashed line). *Units are mm and mm.s$^{-1}$*

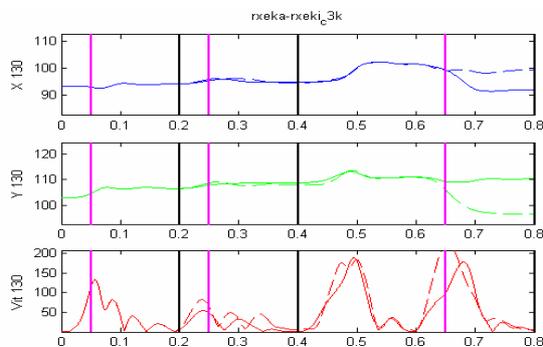

*Figure 4*: Öhman-like model (see Figure 3 for details)